\documentclass[runningheads]{llncs}

\usepackage{graphicx}
\usepackage[cmex10]{amsmath}
\usepackage{subcaption}
\usepackage{hyperref}
\usepackage{cleveref}
\usepackage{rotating}
\usepackage{makecell}
\usepackage{float}

\begin{document}
\title{Global Planar Convolutions for improved context aggregation in Brain Tumor Segmentation}
\titlerunning{Global Planar Convolutions in Brain Tumor Segmentation}
%
\author{
Santi Puch\inst{1} \and
Irina S\'anchez\inst{1} \and
Aura Hern\'andez\inst{2} \and
Gemma Piella\inst{3} \and
Vesna Pr\u{c}kovska\inst{1}}
\authorrunning{S. Puch et al.}
%
\institute{QMENTA, Boston MA \\ 
\email{\{santi,irina,vesna\}@qmenta.com} \and
Computer Vision Center, Universitat Aut\`onoma de Barcelona, Barcelona, Spain \\
\email{aura@cvc.uab.es} \and
SIMBIOsys, Universitat Pompeu Fabra, Barcelona, Spain \\
\email{gemma.piella@upf.edu}}
\maketitle              
\begin{abstract}

In this work, we introduce the Global Planar Convolution module as a building-block for fully-convolutional networks that aggregates global information and, therefore, enhances the context perception capabilities of segmentation networks in the context of brain tumor segmentation. We implement two baseline architectures (3D UNet and a residual version of 3D UNet, ResUNet) and present a novel architecture based on these two architectures, ContextNet, that includes the proposed Global Planar Convolution module. We show that the addition of such module eliminates the need of building networks with several representation levels, which tend to be over-parametrized and to showcase slow rates of convergence. Furthermore, we provide a visual demonstration of the behavior of GPC modules via visualization of intermediate representations. We finally participate in the 2018 edition of the BraTS challenge with our best performing models, that are based on ContextNet, and report the evaluation scores on the validation and the test sets of the challenge.

\keywords{Brain Tumors \and 3D fully-convolutional CNN \and Magnetic Resonance Imaging \and Global Planar Convolution}
\end{abstract}

\section{Introduction}

It is estimated that, as of today, 700.000 people in the United States are living with a primary brain tumor, from which 80\% are benign and 20\% are malignant tumors\cite{tumor_types_nbts}. 
Of all malignant brain tumors, 81\% are gliomas, which are tumors that originate in glial cells\cite{epidemiology_glioma}. Glioblastomas are the most common type of glioma, representing 45\% of all gliomas; they are one of the most aggressive types of brain tumors, having an estimated 5-year relative survival rate of approximately 5\%, which means that only 5\% of people diagnosed with a glioblastoma will still be alive 5 years after being diagnosed \cite{epidemiology_glioma}.

It is clear that such dismal prognosis requires proper treatment planning and follow-up, which can be greatly improved if proper in-vivo, non-invasive delineation and identification of glioma structures is in place. However, this poses a significant burden on the radiologist: multiple imaging modalities have to be assessed in parallel, as each highlights different regions of the tumor, and the process of delineation in a 3D acquisition is tedious and prone to errors. As a consequence, inter-observer variability has been reported to be a major --- if not the largest --- factor of inaccuracy in radiation therapy, constituting the weakest link in the radiotherapy chain that goes from diagnosis and consultation, going through 3D imaging and target volume delineation, to treatment delivery \cite{tumor_delineation}.

Therefore, automating the delineation and identification process on MR images would accelerate treatment planning and improve treatment follow-up. However, the problem of tumor segmentation poses several challenges, such as blurry or smoothed boundaries, variability of shape, location and extension or heterogeneity of appearance of brain tumors on MR images.

The research community has concentrated efforts in order to address the brain tumor segmentation task, and to this end initiatives like the Brain Tumor Segmentation (BraTS) challenge\cite{brats_reference_paper} have made the problem accessible to a larger audience. As a result, a large variety of computational methods have been proposed to automate the delineation of brain tumors. These methods can be broadly categorized in two groups: generative models, which rely on prior knowledge about tissue appearance and distribution; and discriminative models, which directly learn the relationship between the image features and the segmentation labels. Deep Learning approaches, especially Convolutional Neural Networks (CNNs), cover a large portion of the recently proposed discriminative methods. The majority of these works have based their methods in well-known semantic segmentation networks, either using a 2D variant on one or more planes of the brain, or implementing a 3D architecture that takes spatial information in all directions into account. In all these works, several training strategies are leveraged, such as dense training with patches combined with patch sampling schemes, and a distinction between local, refined features and global, coarse features is accomplished via multiresolution approaches or skip connections. 

In this work, we introduce the Global Planar Convolution (GPC) module, a fully-convolutional module that enhances the context perception capabilities of segmentation networks in the context of brain tumor segmentation. We first explore different 3D fully-convolutional architectures for brain tumor segmentation, starting with a 3D variation of UNet \cite{ronnenberg_2015}. We then introduce a variation of such network that incorporates residual elements from \cite{he_2015}, that we call ResUNet. We finally refine this architecture by adding GPC modules; we refer to this network as ContextNet. 
We train these architectures on the 2018 BraTS Challenge dataset, that consists of 210 High Grade Glioma (HGG) cases and 75 Low Grade Glioma (LGG) cases with four MR image modalities and manual annotations of the distinct intra-tumoral structures of interest \cite{brats_reference_paper} \cite{brats_reference_2} \cite{brats_reference_3} \cite{brats_reference_4}. We compare the behavior of the three proposed networks when trained with all the image modalities and when trained only with a subset of them. Then, we show that the addition of Global Planar Convolution modules eliminates the need of building networks with several representation levels (i.e. the set of operations executed at the same spatial resolution level), which are prone to over-parametrization and slow convergence rates. We include a visually guided interpretation of the behavior of GPC modules via visualization of intermediate representations of the network. We finally report the performance of our best performing model ---based on ContextNet--- and a model ensemble on the BraTS validation set. This last model ensemble is submitted to participate in the 2018 edition of the BraTS challenge.

\section{Methods}

\subsection{Data}

The data used in this project originates in the 2018 version of the yearly Multimodal Brain Tumor Segmentation Challenge dataset. This dataset consists of 285 multi-institutional clinically-acquired pre-operative scans. Each multimodal scan is formed by T1, T1-Gd, T2 and T2-FLAIR volumes acquired with various scanners from 19 institutions. All the scans have been segmented manually by one to four raters, and approved by experienced neuroradiologists. The ground-truth labels comprise the enhancing tumor, the peritumoral edema and the necrotic and non-enhancing tumor when present. Each multi-modal scan in the BraTS challenge dataset is co-registered to the same anatomical template, skull-stripped and resampled to 1mm\textsuperscript{3} isotropic resolution. Therefore, no further preprocessing is needed.

\subsection{CNN architectures}
All the Convolutional Architectures implemented and trained are 3D and fully-convolutional by design, meaning that they can be trained using 3D patches of data, and then they can be used for inference on whole brain volumes.

\subsubsection{UNet}
This reference architecture proposed initially in \cite{ronnenberg_2015} is based on \cite{long_2015}, with a similar contracting path, but in order to improve the localization capabilities of the network a supplementary expanding path is introduced, in which high-resolution features from the contracting path are combined with upsampled feature maps. This architecture is extended to 3D by replacing all the convolutions, transposed convolutions and max pooling operations by their 3D alternatives, similarly to \cite{cicek_2016}.

In our experiments we use Rectified Linear Unit (ReLU) as activation function and Batch Normalization \cite{ioffe_2015} after convolutions and before activations.

\subsubsection{ResUNet}

The work of \cite{he_2015} introduced the concept of deep residual learning: instead of stacking a series of layers and letting them learn the desired underlying mapping, these layers can be set to explicitly fit a residual mapping. This residual learning framework not only improved the performance on the image classification task by a wide margin, but also alleviated the degradation problem found when an excessive amount of layers was used.

The motivation of introducing residual layers originates from the fact that the contracting path has a twofold purpose: on one hand it learns increasingly abstract features that encode contextual information necessary to decide \textit{what}, and on the other hand it connects feature maps from lower-level representations with the expanding path to decide \textit{where}. Residual elements allow for an increased number of layers, which has been shown empirically to increase the representational power of the network\cite{szegedy_2015} \cite{he_2015}, thus helping with the first task (\textit{identification}). They also facilitate learning of identity mappings, which enables the possibility of passing low-level representations throughout the network, thus easing the second task (\textit{delineation}).

The architecture of ResUNet is essentially an extension of UNet with residual elements. The convolutional layers in UNet are replaced by residual layers; concretely, 2 residual layers are used at each resolution level. 

Again, as in UNet, we use Rectified Linear Unit (ReLU) as activation function and Batch Normalization after convolutions and before activations.

\subsubsection{ContextNet}
\label{sec:contextnet}

\begin{sidewaysfigure}[!]
    \centering
    \begin{subfigure}[b]{1\textwidth}
        \includegraphics[width=\textwidth]{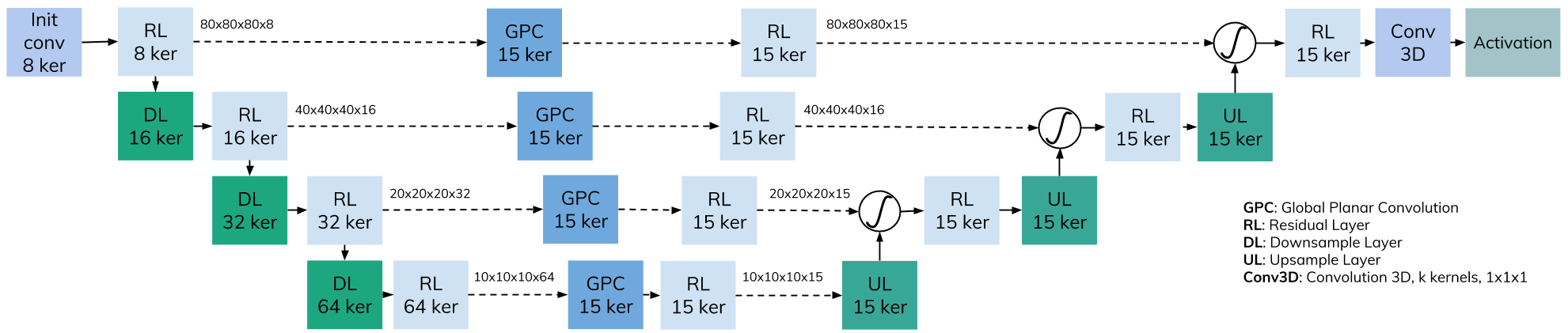}
        \caption{ContextNet architecture overview. Tensor dimensions are specified for a single example during the training phase.}
    \end{subfigure}
    \begin{subfigure}[b]{0.3\textwidth}
        \includegraphics[width=\textwidth]{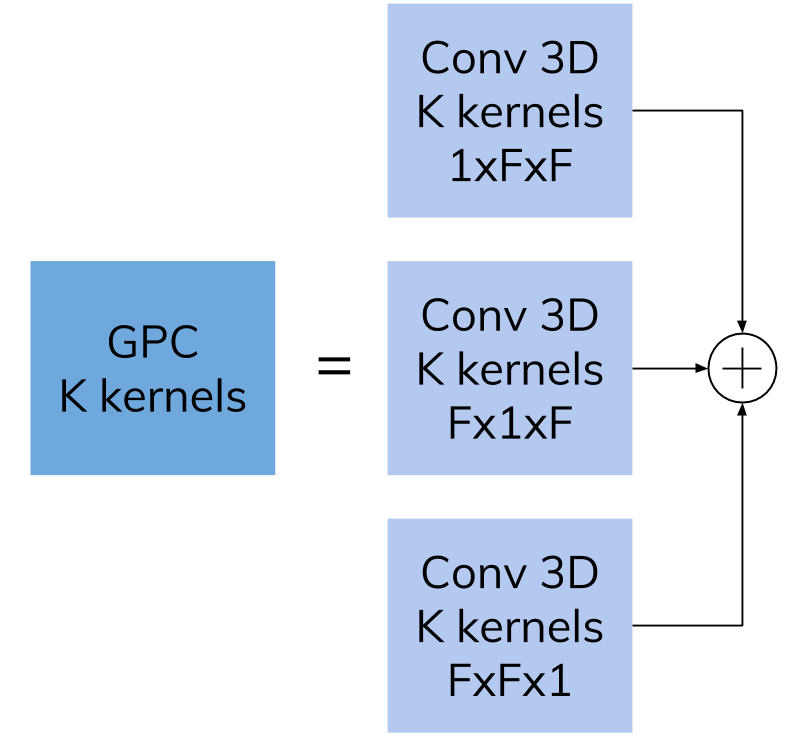}
        \caption{Global Planar Convolution (GPC) module}
        \label{fig:contextnet_modules}
    \end{subfigure}
    \caption{ContextNet architecture diagrams}
    \label{fig:contextnet_diagram}
\end{sidewaysfigure}

ContextNet is a novel architecture introduced in this work that aims to enhance the context-awareness capabilities of 3D imaging segmentation architectures. It is build upon the aforementioned ResUNet architecture and includes the proposed Global Planar Convolution modules, inspired by \cite{peng_2017}. An overview of the architecture is shown in \cref{fig:contextnet_diagram}.

The localization aspect of semantic segmentation networks is addressed by skip-connections and residual elements, as these components let low level representations pass through the network and inform the latest layers about fine-grained spatial details. However, the classification aspect of semantic segmentation networks, that deals with proper identification of the delineated structures, is hindered by the fact that these networks are focused on proper boundary alignment. State-of-the-art classification architectures rely on layers that are globally connected, which in the most extreme case (all the nodes are connected with each other) corresponds to a fully-connected layer. It is clear that such type of operation is not feasible in a fully-convolutional architecture, however we can approximate global connectivity by increasing kernel size in convolutions: in the limit, the kernel is as big as the input feature map, which can be interpreted again as a fully-connected layer. The problem with increasing the kernel size is the computational and memory requirements associated with it, and it is not feasible in the case of 3D CNNs with current accelerated computing hardware. 

However, global connectivity can be approximated by constraining the kernel parameters' subspace. Specifically, we can constrain the convolutional kernels to have one dimension less than they would normally have, which in practice is implemented by having kernels with size 1 in one of the dimensions. This reduction of parameters in one of the three dimensions allows the growth of kernel sizes in the other dimensions, thus providing improved global connectivity.

On the basis of this reasoning, we introduce in this work a new module named \textit{Global Planar Convolution}, abbreviated GPC. A GPC convolves planar filters (i.e. filters in which one of the three dimensions has size 1) in each of the three orthogonal directions, and then combines the resulting planar feature maps via summation. We introduce these modules in between skip-connections, similarly to bottleneck modules in \cite{he_2015}. We further improve the localization capabilities by including an extra residual layer after each GPC module, however these residual layers do not include an activation in the end. The resulting feature maps from each of this altered skip-connections are then summed with the feature maps outputted by the upsampling layers in the expanding path, and the summed feature maps are then passed through an Exponential Linear Unit (ELU) \cite{clevert_2015}. In our experiments we set the filter size of GPC modules to 15.

\subsection{Experimental design}

\subsubsection{Local dataset split}

In order to evaluate the segmentation performance of the proposed architectures, the dataset is split into train and validation sets. We use a split ratio of 70\%-30\% for training and validation, respectively. This results in 199 subjects for training and 86 subjects for evaluation on the tumor segmentation task. As we do not perform any hyper-parameter search procedure, we use the validation set to evaluate if the model is behaving and converging as expected during training, as well as to compare the performance among different models. Thus, we eliminate the need of an additional test set.

\subsubsection{CNN training details}
We use categorical cross-entropy as the loss function to be minimized during training. The complete loss function includes L1 and L2 penalization of the weights (for regularization purposes), with penalization ratios of 1E-6 and 1E-4, respectively. 

All the models are trained using the ADAM optimizer \cite{kingma_2014}. The initial learning is set to 1E-3 in all experiments, and a learning rate decay policy is integrated in order to stabilize training as the training procedure progressed. Concretely, we use an exponential decay of the learning rate every 1000 training steps with a decay rate of 0.9. The number of training steps is set to 35000. The training procedure alternates 1000 training steps with 1 complete evaluation of the model. Batch size is set to 6, which maximizes the memory consumption in the most memory demanding architectures.

During training, the data ingestion pipeline is configured to extract patches of size $80 \times 80 \times 80$ with 50\% probability of being centered on a background voxel and 50\% on a tumor voxel (50\% background, 20\% edema, 15\% enhancing tumor and 15\% necrosis and non-enhancing tumor). Whole brain volumes are used during evaluation in order to provide a realistic value of performance in a real-world scenario. The CNNs are trained on two different hardware configurations, depending on the availability of computing resources: 1) AWS p2.xlarge instance with a single NVIDIA K80 with 12 GiB of GPU memory; 2) on-premises server with two NVIDIA GeFore GTX 1080 Ti with 11GiB of GPU memory.

\subsubsection{Restriction of availability of input modalities}
We perform data ablation experiments by restricting the available input modalities at training time, but always maintaining the minimum required modalities to properly identify all structures, namely T1-Gd and FLAIR. The motivation for such experiments is twofold. First, we want to assess the relative contribution of each modality to the overall segmentation, and inspect if some modalities are redundant or indeed provide useful information. Second, it is convenient and even necessary to have models that can work with a restricted number of modalities (as in some clinical cases not all MR sequences are included in the protocol) even if such models with restricted input information do not perform as well as models trained without data restrictions.

\subsubsection{Restriction of number of representation levels}
\label{sec:restriction_representation_levels}
We hypothesize that the inclusion of GPC modules enables the network to perceive greater context without the need of having several representation levels. By representation levels we mean the set of operations and feature maps that operate at the same spatial resolution. In order to validate this hypothesis, we train two variations of the ContextNet architecture with as little as 2 or 3 representation levels. The first model, with just 2 representation levels, has 32 and 64 kernels at each convolutional layer before the GPC modules, while the GPC modules still have 15 kernels with the same kernel size as in \cref{sec:contextnet}. The second model has 3 representation levels with 16, 32 and 64 kernels at each convolutional layer before the GPC modules, maintaining again the same number of kernels and kernel sizes at the GPC modules and subsequent layers.

\subsubsection{Visualization of GPC feature maps}
\label{sec:gpc_visualization_features}
We extract intermediate representations from the residual layers around the GPC modules and from the GPC modules themselves on all the representation-level variations of ContextNet. The intent of such experiment is to provide insight about the behavior of the GPC modules, and to link such behavior to the performance of these models, despite being aware of the limitations of this method for network interpretability.

\subsection{Evaluation}
The primary evaluation score for the segmentation task is the S{\o}rensen-Dice coefficient, usually abbreviated as DICE. In this context, the DICE coefficient compares the similarity between the set of true examples and the set of positive examples:
\begin{equation}
    DICE = \frac{2 TP}{2 TP + FP + FN}
\end{equation}

The Hausdorff distance is used to evaluate the distance between segmentation boundaries. Results are reported using the 95\% quantile of the maximal surface distance between the ground truth $ P_1 $ and the predicted segmentation $ T_1 $ \cite{menze2015multimodal}:
\begin{equation}
    Haus(P, T) = max({\sup \limits_{p\in\partial P_1} \inf \limits_{t\in\partial T_1} d(p,t), \sup \limits_{t\in\partial T_1} \inf \limits_{p\in\partial P_1} d(t,p)})
\end{equation}

The targets of these evaluation scores are the following tumoral structures:
\begin{itemize}
    \item \textbf{Whole Tumor}: comprises all tumoral structures, i.e. edema, enhancing tumor, non-enhancing tumor and necrosis.
    \item \textbf{Enhancing Tumor}: comprises only the enhancing tumor class.
    \item \textbf{Tumor Core}: encompasses the enhancing tumor, necrosis and non-enhancing tumor, thus excluding edema.
\end{itemize}

\section{Results}

\Cref{tab:dice_local} shows the DICE coefficients on the local validation set for all target structures of all baseline architectures, trained with different data configurations. The best model for whole tumor segmentation is ContextNet trained with all modalities (0.897 DICE score), while the best model for enhancing tumor and tumor core segmentation is ResUNet trained only with T1-Gd, FLAIR and T1, achieving 0.752 and 0.799 DICE scores, respectively. It is particularly remarkable that excluding the T2 from training enables the ResUNet model to better segment the tumor core structures; such behavior can be explained if we consider that the network is encouraged to focus more on structures more noticeable on T1-related modalities (enhancing tumor and tumor core) thanks to the exclusion of redundant information about edema provided by the T2 image (which is clearly visible in FLAIR images).

\begin{table}[H]
    \centering
    \resizebox{\textwidth}{!}{
        \begin{tabular}{l|c|c|c|}
        \cline{2-4}&\textbf{Enhancing Tumor}&\textbf{Whole Tumor} &\textbf{Tumor Core} \\ 
        \hline
        \multicolumn{1}{|l|}{\textbf{UNet - all modalities}} & 0.698 $\pm$ 0.229 & 0.847 $\pm$ 0.095  & 0.694 $\pm$ 0.235  \\ 
        \hline
        \multicolumn{1}{|l|}{\textbf{ResUNet - all modalities}} & 0.739 $\pm$ 0.207& 0.892 $\pm$ 0.064 & 0.785 $\pm$ 0.200  \\ 
        \hline
        \multicolumn{1}{|l|}{\textbf{ResUNet - T1-Gd, FLAIR, T1}}& \textbf{0.752 $\pm$ 0.193} & 0.882 $\pm$ 0.080  & \textbf{0.799 $\pm$ 0.171}  \\ 
        \hline
        \multicolumn{1}{|l|}{\textbf{ResUNet - T1-Gd, FLAIR}}  & 0.723 $\pm$ 0.218 & 0.884 $\pm$ 0.070 & 0.790 $\pm$ 0.184  \\ 
        \hline
        \multicolumn{1}{|l|}{\textbf{ContextNet - all}} & \textbf{0.752 $\pm$ 0.207} & \textbf{0.897 $\pm$ 0.059} & 0.797 $\pm$ 0.195  \\ 
        \hline
        \multicolumn{1}{|l|}{\textbf{ContextNet - T1-Gd, FLAIR, T1}} & 0.743 $\pm$ 0.216 & 0.881 $\pm$ 0.071  & 0.770 $\pm$ 0.211  \\ 
        \hline
        \multicolumn{1}{|l|}{\textbf{ContextNet - T1-Gd, FLAIR}}& 0.734 $\pm$ 0.231 & 0.878 $\pm$ 0.080 & 0.770 $\pm$ 0.216  \\ 
        \hline
        \end{tabular}
    }
    \caption{DICE coefficients (avg $\pm$ std) of baseline architectures trained with different data configurations. Scores are computed on the local validation set.}
    \label{tab:dice_local}
\end{table}

We show in \cref{tab:dice_rl_local} a comparison of the performance (according to the DICE score) of the different representation level variations of ContextNet, as detailed in \cref{sec:restriction_representation_levels}. We include the scores of ResUNet trained with all modalities, as it is the most valid non-GPC model to be used for comparison. Both ContextNet models with reduced number of representation levels match or even surpass the performance of the ResUNet model at whole tumor and enhancing tumor segmentation. We hypothesize that GPC modules enable the aggregation of contextual information without the need of obtaining a deep representation via several pooling operations, which permits the proper segmentation of large structures such as the whole tumor. At the same time, by reducing the number of coarse features associated with increased representation levels the network can focus on fine details, such as the enhancing tumor. Furthermore, a reduction of representation levels entails a reduction of the number of trainable parameters, as can be seen in \cref{tab:dice_rl_local}. The downside is that the depth of the network is severely reduced, and much more complex structures such as the necrotic and non-enhancing tissue are harder to segment for such networks.

\begin{table}[h!]
    \centering
    \resizebox{\textwidth}{!}{
        \begin{tabular}{l|c|c|c|c|}
        
        \cline{2-5} & \textbf{Enhancing Tumor} & \textbf{Whole Tumor} & \textbf{Tumor Core} & \textbf{\# parameters} \\
        \hline
        \multicolumn{1}{|l|}{\textbf{ResUNet}} & 0.739 $\pm$ 0.207& 0.892 $\pm$ 0.064 & 0.785 $\pm$ 0.200 & 2M \\ 
        \hline
        \multicolumn{1}{|l|}{\textbf{ContextNet}} & \textbf{0.752 $\pm$ 0.207} & \textbf{0.897 $\pm$ 0.059} & \textbf{0.797 $\pm$ 0.195} & 1.7M \\ 
        \hline
        \multicolumn{1}{|l|}{\textbf{ContextNet - 3 RL}} & 0.746 $\pm$ 0.217 & 0.894 $\pm$ 0.062 & 0.763 $\pm$ 0.225 & 1.6M \\ 
        \hline
        \multicolumn{1}{|l|}{\textbf{ContextNet - 2 RL}} & 0.751 $\pm$ 0.208 & 0.892 $\pm$ 0.064 & 0.750 $\pm$ 0.229 & 1.3M \\ 
        \hline
        
        \end{tabular}
    }
    \caption{DICE coefficients (avg $\pm$ std) of baseline ResUNet and variations of ContextNet with different number of representation levels. All MRI modalities are used to train these networks. Scores are computed on the local validation set.}
    \label{tab:dice_rl_local}
\end{table}

\begin{figure}[H]
    \centering
    \includegraphics[width=\textwidth]{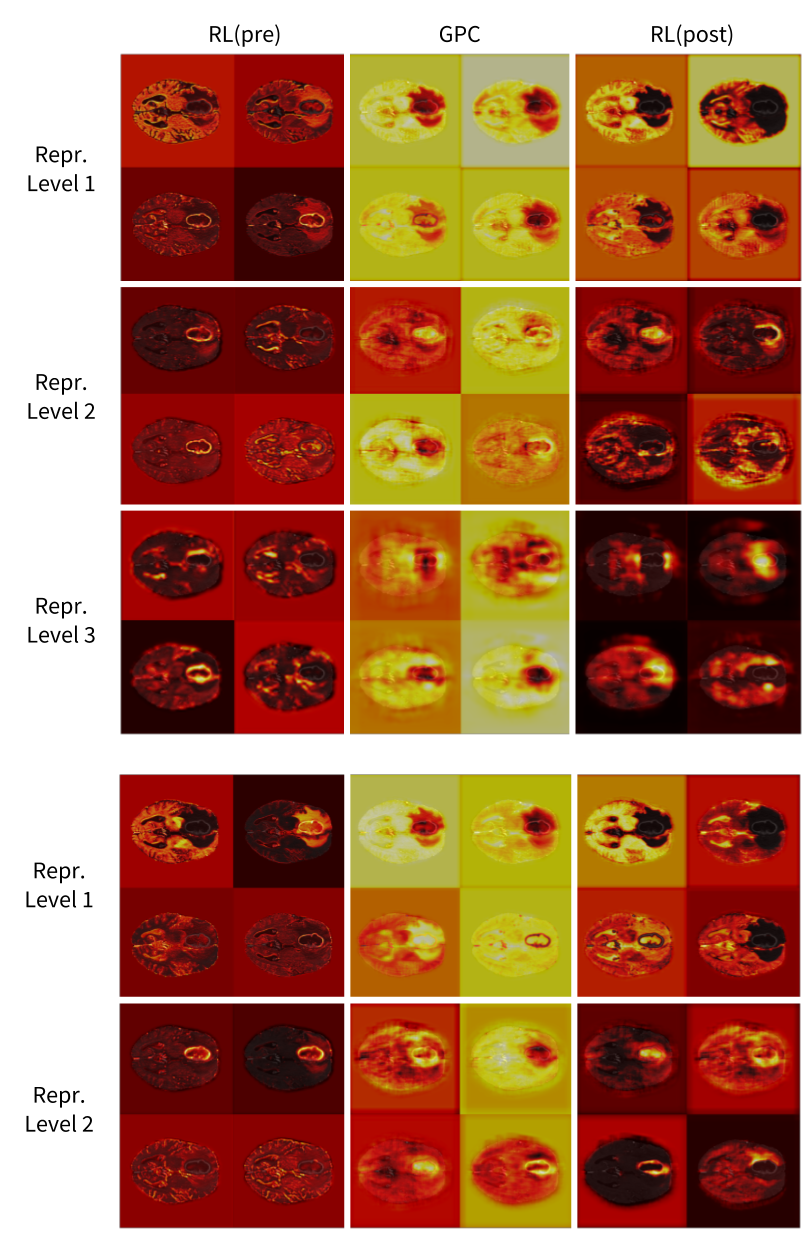}
    \caption{Feature map visualization for GPC module interpretability. Top figure shows the activations of the ContextNet variant with 3 representation levels, while the bottom figure shows the activations of the ContextNet variant with 2 representation levels.}
    \label{fig:activations_lr_2_3}
\end{figure}

\Cref{fig:activations_lr_2_3} depicts the feature maps extracted from the residual layers and GPC modules on the ContextNet models with reduced representation levels, as explained in \cref{sec:gpc_visualization_features}. We show only the first 4 feature maps with the highest mean absolute value at each representation level and stage, and we organize the information in a grid in which each row corresponds to a representation level and each column corresponds to a module type in the network (pre-GPC residual layer, GPC, and post-GPC residual layer).

Overall it is clear that the deeper the representation level, the coarser the features the network is able to extract. It can also be seen that, as we move from the pre-GPC residual layer to the GPC module and then the post-GPC residual layer, the features that the network extracts are increasingly abstract, even at the first representation levels. For instance, on the first level of both models the pre-GPC residual layer is enhancing fine-details of the images, such as the enhancing tumor ring; then the GPC module aggregates contextual information and captures global features such as the healthy part of the brain (the activation around whole tumor region is close to 0); finally the post-GPC residual layer combines the global features extracted by the GPC modules and refines their boundaries. 

\begin{figure}[H]
    \centering
    \includegraphics[width=\textwidth]{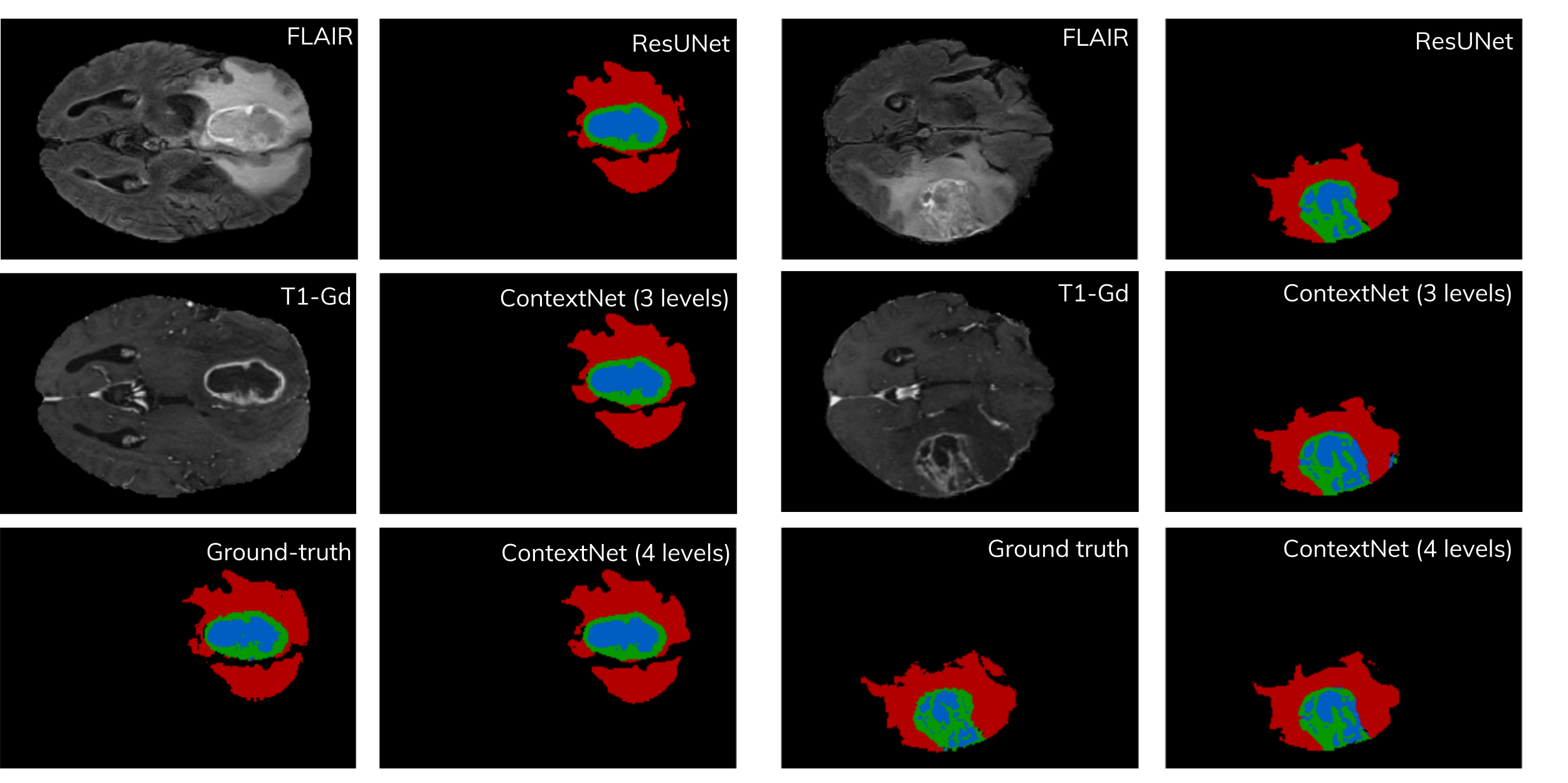}
    \caption{From top to bottom, left to right: FLAIR and T1-Gd MR modalities, ground-truth labels and segmentations produced by ResUNet, ContextNet trained with 3 representation levels and ContextNet with all (4) representation levels of 2 subjects from the local validation split of the BraTS 2018 dataset.}
    \label{fig:brats_qualitative_examples}
\end{figure}

Predictions obtained from ResUNet, ContextNet constrained to three representation levels and full ContextNet are shown in \cref{fig:brats_qualitative_examples}. On one hand, the ContextNet variant with three representation levels is shown to perform similarly to other networks on the first subject. On the other hand, it can be noticed that the depth reduction in this ContextNet variant affects the segmentation of the necrotic tissue on the second subject, in which both the full ContextNet and ResUnet are able to produce a more refined representation than the limited representation-level ContextNet.

Finally we evaluate the best performing ContextNet and ResUNet models and an ensemble of these two models on the BraTS 2018 validation data. As discussed in \cite{kamnitsas_2017}, model ensembling yields more robust segmentation maps by reducing the influence of the hyper-parameters and configurations of individual models. Specifically, we compute the average confidence score per class for each voxel across the models in the ensemble, and we obtain the final segmentation by assigning to each voxel the class with the highest average confidence score. As a consequence of model ensembling, we observe improved DICE scores and Hausdorff 95\% quantile distances in practically all structures (shown in \cref{tab:dice_brats_validation}). Therefore, we submit this model ensemble to the BraTS 2018 challenge\cite{brats_2018_reference_manuscript} and we report the resulting scores on the test set in \cref{tab:dice_brats_test}. 

\begin{table}[h!]
    \centering
    \resizebox{\textwidth}{!}{
        \begin{tabular}{l|c|c|c|c|c|c|}
            \cline{2-7}
            \multicolumn{1}{c|}{} & \multicolumn{3}{c|}{\textbf{Dice}} & \multicolumn{3}{c|}{\textbf{Hausdorff 95}} \\ 
            \cline{2-7} 
            \multicolumn{1}{c|}{} & \textit{ET} & \textit{WT} & \textit{TC} & \textit{ET} & \textit{WT} & \textit{TC} \\  
            \cline{1-7}
            \multicolumn{1}{|l|}{\textbf{ResUNet}}     & 0.729 $\pm$ 0.279    & 0.882 $\pm$ 0.071  & 0.741 $\pm$ 0.256  & 5.578 $\pm$ 11.249   & 9.896 $\pm$ 16.803  & 9.532 $\pm$ 12.407   \\ \hline
            \cline{2-7} 
            \multicolumn{1}{|l|}{\textbf{ContextNet}}   & 0.735 $\pm$ 0.281    & 0.883  $\pm$ 0.112 & 0.753  $\pm$ 0.269  & 7.004  $\pm$ 13.944  & \textbf{7.594  $\pm$ 12.453}   & 9.505  $\pm$ 11.557 \\ \hline
            \cline{2-7}
            \multicolumn{1}{|l|}{\textbf{Ensemble}}       & \textbf{0.758 $\pm$ 0.264}    & \textbf{0.895 $\pm$ 0.07}  & \textbf{0.774 $\pm$ 0.253}  & \textbf{4.502 $\pm$ 8.227}   & 10.656 $\pm$ 19.286 & \textbf{7.103 $\pm$ 7.084} \\ \hline
        \end{tabular}
    }
    \caption{Evaluation scores (avg $\pm$ std) obtained on the BraTS 2018 validation set. The evaluated models are ContextNet trained on all MR image modalities, ResUNet trained on T1-Gd, T1 and FLAIR, and an ensemble of both models.}
    \label{tab:dice_brats_validation}
\end{table}

\begin{table}[h!]
    \centering
    \resizebox{\textwidth}{!}{
        \begin{tabular}{l|c|c|c|c|c|c|}
            \cline{2-7}
            \multicolumn{1}{c|}{} & \multicolumn{3}{c|}{\textbf{Dice}} & \multicolumn{3}{c|}{\textbf{Hausdorff 95}} \\ 
            \cline{2-7} 
            \multicolumn{1}{c|}{} & \textit{ET} & \textit{WT} & \textit{TC} & \textit{ET} & \textit{WT} & \textit{TC} \\  
            \cline{1-7}
            \multicolumn{1}{|l|}{\textbf{Ensemble}} & 0.694 $\pm$ 0.289 & 0.856 $\pm$ 0.147 & 0.754 $\pm$ 0.283 & 6.872 $\pm$ 13.21 & 9.676 $\pm$ 15.947 & 8.123 $\pm$ 12.713 \\ \hline
        \end{tabular}
    }
    \caption{Evaluation scores (avg $\pm$ std) of the submitted segmentation model on the BraTS 2018 test set.}
    \label{tab:dice_brats_test}
\end{table}

\section{Conclusion}

In this work, we present several 3D fully-convolutional CNNs to address the task of automatic tumor segmentation from magnetic resonance images, with the objective of accelerating and improving radiotherapy planning and monitoring of patients with gliomas of varied grades. We start with a baseline architecture (UNet) and gradually improve its performance by adding residual elements (ResUNet) and enlarging the receptive field of its components via GPC modules (ContextNet).

We further investigate the behavior of the GPC modules by training networks with a limited number of representation levels and visualizing their intermediate representations, and show that equivalent performance can be achieved using GPC modules even when the number of representation levels (and consequently the depth and number of trainable parameters) of the network is considerably reduced. 

Future work includes improving the performance of individual models by means of hyper-parameter optimization, uncertainty estimation via Monte-Carlo Dropout or related techniques, in-depth investigation of intermediate representations and use of other deep learning interpretability methods to better understand the behavior of the proposed GPC modules.

%
%
%

\bibliographystyle{splncs04}
\bibliography{references}

\end{document}